\documentclass[twocolumn,amsmath,aps]{revtex4}
\usepackage{graphicx,color}
\usepackage{CJK}
\usepackage{bm}
\usepackage[hypertex]{hyperref}

\newcommand{\be}{\begin{equation}}
\newcommand{\ee}{\end{equation}}
\newcommand{\bea}{\begin{eqnarray}}
\newcommand{\eea}{\end{eqnarray}}
\newcommand{\bsube}{\begin{subequations}}
\newcommand{\esube}{\end{subequations}}

\newcommand{\Eq}[1]{Eq.\,(\ref{#1})}
\newcommand{\Eqs}[1]{Eqs.\,(\ref{#1})}

\newcommand{\la}{\langle}
\newcommand{\ra}{\rangle}

\newcommand{\nl}{\nonumber \\}

\newcommand{\gam}{\gamma}

\newcommand{\beq}{\begin{equation}}
\newcommand{\eeq}{\end{equation}}
\newcommand{\beqn}{\begin{eqnarray}}
\newcommand{\eeqn}{\end{eqnarray}}
\newcommand{\bsub}{\begin{subequations}}
\newcommand{\esub}{\end{subequations}}

\newcommand{\fdg}{f^\dagger}

\begin{document}
\begin{CJK*}{GBK}{Song}

\title{Cross correlation mediated by distant Majorana zero modes with no overlap}

\author{Lupei Qin}
\affiliation{Center for Joint Quantum Studies and Department of Physics,
School of Science, Tianjin University, Tianjin 300072, China}

\author{Wei Feng}
\affiliation{Center for Joint Quantum Studies and Department of Physics,
School of Science, Tianjin University, Tianjin 300072, China}

\author{Xin-Qi Li}
\email{xinqi.li@tju.edu.cn}
\affiliation{Center for Joint Quantum Studies and Department of Physics,
School of Science, Tianjin University, Tianjin 300072, China}

\date{\today}

%% \maketitle
\begin{abstract}
Existing studies via shot noise calculation conclude that
the cross correlation between the currents in the two leads
connected by a pair of Majorana zero modes (MZMs)
vanishes when their coupling energy $\epsilon_M\to 0$.
Motivated by the intrinsic nature of nonlocality of the MZMs,
we revisit this important problem
and propose an experimental scheme to demonstrate
the nonvanishing cross correlation even at the limit $\epsilon_M\to 0$.
The proposed scheme employs
the Andreev-process-associated branch circuit currents,
which are theoretically obtained by applying
a decomposition analysis for the total currents
while in practical measurement, are accessible directly.
For different bias voltage setup,
we find intriguing results of both negative and positive correlations
and carry out simple physical understanding using a quantum jump technique.
Importantly,
combining together with the evidence of the zero-bias-peak of conductance,
the nonlocal cross correlation predicted in this work
can help to definitely confirm the existence of the nonlocal MZMs.
\end{abstract}

% \pacs{03.65.Yz,03.65.Sq,31.15.xv,31.15.xg}

\maketitle

\section{Introduction}

The nonlocal nature of the Majorana zero modes (MZMs) and the obeying
non-Abelian braiding statistics promise a sound potential
for topological quantum computation \cite{Kita01,Kit03,Sar08,Sar15}.
Throughout the past decade, the search for reliable MZMs has become a major theme
in condensed matter physics \cite{Kou12,Kou19,Ali12,Flen12,Bee13,Agu17a,Ore18}.
In order to identify the Majorana zero modes (MZMs),
rich transport phenomena have been proposed,
such as the fractional Josephson effects \cite{Sar10,Opp10,Fu09,Cay17},
peculiar noise behaviors \cite{Dem07,Bee08,Zoch13,Li12,Law09,BNK11,Law13,Ore15,Has20},
and the famous Majorana quantized conductance $2e^2/h$
\cite{Law09,BNK11,Sarma01,Flen10,Flen16}.
Recent interests also include the nonlocal transport conductances
\cite{Gla16,DS17d,Mar18b,Mar18c,Sch-1,Sch-2,BCS-1a,BCS-1,BCS-2}
and the proposals to distinguish the nonlocal MZMs
from the topologically trivial Andreev states \cite{DS17,Agu17,Cay19,Agu18,Vu18}.

The most direct evidences of the genuinely nonlocal nature of the MZMs
should be the Majorana teleportation \cite{Sem06ab,Lee08,Fu10}
and/or the cross correlation of two remote majoranas
($\gamma_1$ and $\gamma_2$) \cite{Dem07,Bee08,Zoch13,Li12,Law09}.
Both phenomena of Majorana teleportation and nonlocal cross correlation
are rather transparent using the low energy effective Hamiltonian description,
especially within the framework of ``second quantization" in terms of
the occupation-number-states $|0\ra$ and $|1\ra$
of the MZMs associated regular fermion (the $f$ quasiparticle).
Consider injecting an electron from outside into the superconductor
through, for instance, the left MZM $\gamma_1$.
The occupation of the nonlocal $f$ quasiparticle should allow for
extracting a particle (either an electron or a hole)
through the other distant right-side MZM $\gamma_2$,
thus expecting the so-called teleportation phenomenon.
The same reason allows us to expect nonlocal cross correlation
between the currents in the distant leads.
More specifically, consider a two-lead (three-terminal) setup
\cite{Dem07,Bee08,Zoch13,Sch-1,Sch-2,BCS-1a,BCS-1,BCS-2}.
Can we find nonlocal cross correlation between the currents in different leads?
A few studies showed that the cross correlation vanishes
at the limit $\epsilon_M\to 0$ \cite{Dem07,Bee08,Zoch13}.

\begin{figure}[h]
\includegraphics[scale=0.5]{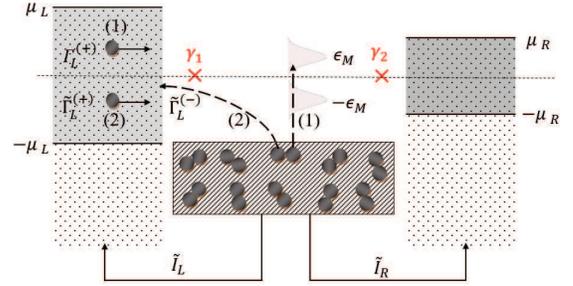}
\caption{
Schematic circuit diagram for the Andreev reflection (AR) and inverse AR processes
in the two-lead (three-terminal) device under consideration.
In the AR process, taking the left lead as example, two electrons in the lead
enter the superconductor, successively and coherently, and form a Cooper pair.
The two electrons are of energy in resonance with
(have Lorentzian centers at) $\epsilon_M$ and $-\epsilon_M$, respectively,
as seen in \Eqs{rates} and (\ref{t-delt}) in the text.
The inverse AR process splits a Cooper pair and generates the excitation
of an $f$ quasiparticle in the superconductor,
meanwhile, sends an electron back into the lead.
In the plot, we only show the left-side current and the local Andreev process,
while the right-side current and the crossed Andreev process
can be similarly understood.
In the diagram we also schematically show the net branch circuit currents
$\widetilde{I}_L$ and $\widetilde{I}_R$, flowing back to the leads.    }
\end{figure}

In this work, applying the master equation approach
(and using the occupation-number-state representation),
we show that it is possible by extracting component currents
from the total lead-currents
to expose the intrinsic nonlocal cross correlation between them,
even at the limit $\epsilon_M\to 0$, owing to the nonlocal nature
of the MZMs associated $f$ quasiparticle.
Moreover, the component currents extracted from the total ones
are nothing but the branch circuit currents
flowing back to the individual leads from the superconductor,
which are measurable in experiment, as schematically shown in Fig.\ 1.
Combining together with the evidence of the
zero-bias-peak of conductance (ZBPC),
this nonlocal cross correlation should allow us to
definitely confirm the existence of the nonlocal MZMs.
This is the most important issue at present stage in the Majorana community.

\section{Master Equation Approach}

For completeness, we briefly outline the Majorana master equation (MME) approach
based on Refs.\ \cite{Li20,Li20b},
which deal with a pair of MZMs
embedded in a two-lead (three-terminal) nonlocal transport setup.
The low-energy effective description for a pair of MZMs
can be commonly formulated by the Hamiltonian
$H_M=i\epsilon_M \gam_1\gam_2$,
where $\epsilon_M$ is the coupling energy of the MZMs $\gamma_1$ and $\gamma_2$.
The Majorana operators are related to the regular complex fermion
through the transformation of $\gam_1=f+\fdg$ and $\gam_2=-i(f-\fdg)$.
Using the complex fermion representation, for a two-lead setup of transport,
the tunnel-coupling of the MZMs to the leads is given by \cite{Li12}
\bea\label{Ham2}
H' = \sum_{\alpha=L,R}\sum_k t_{\alpha k}
\left[ (b^{\dagger}_{\alpha k}f
+(-1)^{\tilde{\alpha}+1} b^{\dagger}_{\alpha k}f^{\dagger})+{\rm H.c.}\right] \,.
\eea
$b^{\dagger}_{\alpha k}$ ($b_{\alpha k}$) are the creation (annihilation) operators
of electrons in the leads, while the leads are described by
$H_{\rm leads}= \sum_{\alpha=L,R}\sum_k \epsilon_{\alpha k} b^{\dagger}_{\alpha k} b_{\alpha k}$.
It should be noted that in $H'$ the tunneling terms only conserve charges modulo $2e$,
which actually correspond to the well known Andreev reflection (AR) process.
In \Eq{Ham2}, the sign factor is determined by assigning $\tilde{\alpha}=1$ and $2$,
corresponding to $\alpha=L$ and $R$, respectively.

The master equation approach to quantum transport
simply follows the theoretical treatment of quantum dissipation,
where the system-of-interest is coupled to an environment
usually via energy exchange, but not involving particle exchange.
For quantum transport, the system-of-interest is the central device,
while the environment is the transport leads. In the transport problem,
particle exchange/tunneling between the leads and the central device
is seemingly causing certain difficulty
to describe the central device using the density matrix.
However, after introducing the occupied and unoccupied probabilities of states,
the fundamental requirement of ${\rm Tr}\rho=1$ is guaranteed
for the density matrix operator $\rho$.
The transport master equation is thus well justified
and has been applied broadly in practice.

Formally, the normal tunneling terms in \Eq{Ham2} (e.g., $b^{\dagger}_{\alpha k}f$)
correspond to the rotating-wave-approximation terms in quantum optics,
while the Andreev process terms (e.g., $b^{\dagger}_{\alpha k}f^{\dagger}$)
correspond to the counter-rotating-wave terms.
In quantum optics, in the presence of the latter type of terms,
the master equation has been well established,
especially under the Born-Markov approximation \cite{WaMi94}.
Following such type of results, and in particular the specific Refs.\ \cite{Li05,Li14},
we constructed the MME as \cite{Li20b,Li20}
\bea\label{ME-1}
\dot{\rho} &=& -\frac{i}{\hbar}[H_M,\rho]
+ \sum_{\alpha=L,R} \left(\Gamma_{\alpha}^{+} {\cal D}[f^{\dagger}]\rho
+ \Gamma_{\alpha}^{-} {\cal D}[f]\rho \right)   \nl
&& + \sum_{\alpha=L,R} \left(\widetilde{\Gamma}_{\alpha}^{+} {\cal D}[f]\rho
+ \widetilde{\Gamma}_{\alpha}^{-} {\cal D}[f^{\dagger}]\rho \right)  \,.
\eea
Here the Lindblad superoperator is defined by
${\cal D}[A]\rho= A\rho A^{\dagger}-\frac{1}{2}\{A^{\dagger}A, \rho\}$
and the tunnel-coupling rates are introduced as
\begin{subequations}\label{rates}
\bea %\label{rate}
&& \Gamma_{\alpha}^{{\pm}}=\Gamma^e_{\alpha} N_{\alpha}^{(\pm)}\,, ~~
N_{\alpha}^{(\pm)}= \int d\omega n_{\alpha}^{(\pm)}(\omega)
\widetilde{\delta}(\omega-\epsilon_M) \,,  \nl
\\
&& \widetilde{\Gamma}_{\alpha}^{{\pm}}=\Gamma^h_{\alpha}
\widetilde{N}_{\alpha}^{(\pm)}\,, ~~
\widetilde{N}_{\alpha}^{(\pm)}= \int d\omega n_{\alpha}^{(\pm)}(\omega)
\widetilde{\delta}(\omega+\epsilon_M) \,. \nl
\eea
\end{subequations}
The superscripts ``$e$" and ``$h$" of the rates denote coupling of the $f$ quasiparticle
to the leads via ``electron" and ``hole" components.
We have also denoted the Fermi occupied function by $n^{(+)}_{\alpha}$
and the empty function by $n^{(-)}_{\alpha}=1-n^{(+)}_{\alpha}$.
The spectral function $\widetilde{\delta}(\omega\mp\epsilon_M)$ is a generalization
of the Dirac $\delta$-function
\bea\label{t-delt}
\widetilde{\delta}(\omega\mp\epsilon_M)
= \frac{1}{\pi}\frac{\Gamma}{(\omega\mp\epsilon_M)^2+\Gamma^2} \,,
\eea
where the broadening is given by
$\Gamma=\sum_{\alpha} (\Gamma^e_{\alpha}+\Gamma^h_{\alpha})/2$.
This type of generalization in terms of the Lorentzian spectral function
properly accounts for the level broadening effect, which directly follows
the spirit of the self-consistent Born approximation \cite{Li14}
and works perfectly well for transport through
single-level systems (such as the single-level quantum dot).
This generalization makes the transport master equation
applicable under small bias voltage,
while it is well known that the usual Born-Markov-Lindblad master equation
is applicable only under large bias limit
(with voltage much larger than the level broadening widths).
Here, in the Majorana case, the subgap $f$ quasiparticle is simply
a single-level system within the low-energy effective description.
The only unique feature is the presence of
the {\it Andreev-process} associated counter-rotating-wave terms,
which results in the two terms in the second round brackets in \Eq{ME-1},
while the two Lindblad terms in the first round brackets in \Eq{ME-1}
describe the {\it normal tunneling process}.
We finally mention that, for the two types of process,
the energy conservation is manifested differently
in the rate expressions, i.e., by the different centers
of the Lorentzian spectral functions $\widetilde{\delta}(\omega\mp \epsilon_M)$,
see also illustration in Fig.\ 1.

\section{Transient Current and Its Decomposition}

The MME can be straightforwardly solved using
the number-state basis $\{|0\ra, |1\ra \}$ of the complex fermion (the $f$ particle).
Let us denote the solution of the density matrix as
$\rho=p_0|0\ra\la 0|+p_1|1\ra\la 1|$.
As an example, the left-lead current, $I_L=I^{(1)}_L+I^{(2)}_L$,
can be calculated as \cite{Li20b}
\bea\label{IL-1}
I^{(1)}_L &=& \frac{e}{\hbar} \left(\Gamma^{+}_L p_0 - \Gamma^{-}_L p_1\right) , \nl
I^{(2)}_L &=& \frac{e}{\hbar} \left(\widetilde{\Gamma}^{+}_L p_1
- \widetilde{\Gamma}^{-}_L p_0\right) \,.
\eea
Physically, $I^{(1)}_L$ is contributed by the conventional tunneling process
and $I^{(2)}_L$ is from the Andreev process, including also the crossed AR (CAR) process.

Noting that $\Gamma^{+}_L+\Gamma^{-}_L=\Gamma^e_L$ and
$\widetilde{\Gamma}^{+}_L+\widetilde{\Gamma}^{-}_L=\Gamma^h_L$,
we reexpress the total left-lead current as
\bea
&& I_L = I^{(1)}_L + I^{(2)}_L  \nl
&& = \frac{e}{\hbar} \left[ \left(\Gamma^{+}_L (1-p_1) - \Gamma^{-}_L p_1\right)
+ \left(\widetilde{\Gamma}^{+}_L p_1 - \widetilde{\Gamma}^{-}_L (1-p_1)\right)  \right] \nl
&& = \frac{e}{\hbar} \left[\Gamma^{+}_L - \widetilde{\Gamma}^{-}_L
- p_1 (\Gamma^e_L-\Gamma^h_L) \right]  \,.
\eea
In the case of $\epsilon_M=0$ or small $\epsilon_M$,
the difference of $\Gamma^e_L$ and $\Gamma^h_L$
is zero or negligibly small.
We thus find that the total current $I_L$ (and also $I_R$) does not depend on time,
even in the transient process with time dependent $p_1(t)$.
%%  //
This unusual feature is rooted in the unique Andreev process involving here.
That is, entering of the ``first" electron is conditioned on
the unoccupied state of the the $f$ quasiparticle ($n_f=0$),
while entering of the ``second" electron is conditioned on its occupied state ($n_f=1$).
As a consequence, the total current does not depend on the occupation probability $p_1$.
We may point out that this conclusion holds under the Markovian approximation,
which guarantees the use of the rate equation
and is valid on timescale longer than
the so-called memory time of the reservoir (transport lead) electrons.

The feature that $I_L$ and $I_R$ are independent of the occupation probability $p_1(t)$
is seemingly raising an obstacle for us to investigate the cross correlation between them,
since the current correlation function simply characterizes
a disturbance of the occupation probability and the subsequent response
associated with two successive current measurements.
That is, the correlation function of the total currents $I_L$ and $I_R$
cannot reveal the intrinsic nature of
the dynamical disturbance-and-response process.
However, as to be clear soon,
we can consider the cross correlation of the partial components of the currents.
To achieve this, below we analyze first
the ``inside structure" of the entire lead current,
which is connected with the total occupation probability of the $f$ quasiparticle.

For the brevity of notation,
let us introduce the $f$ particle's excitation rate
$r_1=\sum_{\alpha=L,R}(\Gamma^{+}_{\alpha}+\widetilde{\Gamma}^{-}_{\alpha})$,
its deexcitation rate
$r_2=\sum_{\alpha=L,R} (\Gamma^{-}_{\alpha}+\widetilde{\Gamma}^{+}_{\alpha})$,
and the total rate
$r_1+r_2=\sum_{\alpha=L,R}(\Gamma^e_{\alpha}+\Gamma^h_{\alpha}) = 2\Gamma$.

\subsection{Unoccupied Initial Condition}

Let us consider first the time-dependent solution under the initial condition
of {\it empty occupation} of the $f$ particle state ($n_f=0$), which reads
\bea\label{p1p0t-1}
p_1 &=& \frac{r_1}{2\Gamma} G(t)  \,, \nl
p_0 &=& \frac{r_2}{2\Gamma} G(t) + [1-G(t)]  \,.
\eea
Here we introduced $G(t)=1-e^{-2\Gamma t}$.
We organize the solution in this form for a particular advantage.
It shows that the probability evolution contains two channels:
one is the gradual formation of the {\it steady-state occupation channel}
associated with $G(t)$ (from 0 to 1);
and another one is the gradual decay of the {\it empty occupation channel}
described by $1-G(t)$ (from 1 to 0).
Of great interest is that we can treat
the evolution of these two channels being independent,
as to be clarified with details in the following.

The steady-state-occupation-channel would lead to the stationary probabilities
$\bar{p}_1=\frac{r_1}{2\Gamma}$ and $\bar{p}_0=\frac{r_2}{2\Gamma}$.
Substituting them into \Eq{IL-1}, we obtain the steady-state currents
$I^{(1)}_{L,st}$ and $I^{(2)}_{L,st}$.
Let us consider first the current $I^{(1)}_{L,st}$,
which can be decomposed into three components as
\bea
I^{(1)}_{L,st}(A_1)&=& \frac{e}{\hbar} \,
\left(\Gamma^+_L \widetilde{\Gamma}^+_L - \Gamma^-_L \widetilde{\Gamma}^-_L \right)/2\Gamma  \,,  \nl
I^{(1)}_{L,st}(A_2)&=& \frac{e}{\hbar} \,
\left(\Gamma^+_L \widetilde{\Gamma}^+_R - \Gamma^-_L \widetilde{\Gamma}^-_R \right)/2\Gamma \,,  \nl
I^{(1)}_{L,st}(A_3)&=& \frac{e}{\hbar} \,
\left(\Gamma^+_L \Gamma^-_R - \Gamma^-_L \Gamma^+_R \right)/2\Gamma    \,.
\eea
Using \Eq{rates}, we further reexpress them as
\bea\label{IL-e1-2a}
I^{(1)}_{L,st}(A_1)&=& \frac{e}{\hbar} \frac{\Gamma^e_L\Gamma^h_L}{2\Gamma}
\left( N^{(+)}_L - \widetilde{N}^{(-)}_L \right)  \,,  \nl
I^{(1)}_{L,st}(A_2)&=& \frac{e}{\hbar} \frac{\Gamma^e_L\Gamma^h_R}{2\Gamma}
\left( N^{(+)}_L - \widetilde{N}^{(-)}_R \right)  \,, \nl
I^{(1)}_{L,st}(A_3)&=& \frac{e}{\hbar} \frac{\Gamma^e_L\Gamma^e_R}{2\Gamma}
\left( N^{(+)}_L - N^{(+)}_R \right)   \,.
\eea
Based on this form, we can easily convert them into the usual results
from the scattering matrix or nonequilibrium Green's function approach \cite{Li20b}
\bea\label{IL-e1-2b}
I^{(1)}_{L,st}(A_1)&=&
\frac{e}{h}\int^{\mu_L}_{-\mu_L} d\omega\, {\cal T}^{eh}_{LLA}(\omega) \,, \nl
I^{(1)}_{L,st}(A_2)&=&
\frac{e}{h}\int^{\mu_L}_{-\mu_R} d\omega\, {\cal T}^{eh}_{LRA}(\omega) \,, \nl
I^{(1)}_{L,st}(A_3)&=&
\frac{e}{h}\int^{\mu_L}_{\mu_R} d\omega\, {\cal T}^{ee}_{LR}(\omega)  \,.
\eea
Here, the three transport coefficients (and the associated currents)
correspond to, respectively,
the local AR at the left lead,
the crossed AR between the two leads,
and the electron transmission from the left to the right lead.
More explicitly, substituting \Eqs{rates} and (\ref{t-delt}) into \Eq{IL-e1-2a},
we identify the transport coefficients as
\bea\label{T-coef-1}
{\cal T}^{eh}_{LLA}(\omega)
&=& \Gamma^e_L\Gamma^h_L / |Z|^2 \,, \nl
{\cal T}^{eh}_{LRA}(\omega)
&=& \Gamma^e_L\Gamma^h_R / |Z|^2  \,, \nl
{\cal T}^{ee}_{LR}(\omega)
&=& \Gamma^e_L\Gamma^e_R / |Z|^2   \,,
\eea
where $|Z|^2=(\omega-\epsilon_M)^2+\Gamma^2$.
Actually, one can check that these coefficients
fall into the well-known form obtained from
the nonequilibrium Green's function formalism
or the $S$ matrix scattering approach, i.e.,
${\cal T}_{\alpha\beta}(\omega)
={\rm Tr}[\Gamma_{\alpha}G^r(\omega)\Gamma_{\beta}G^a(\omega)]$,
where the retarded/advanced Green's function reads as
$G^{r(a)}(\omega)=(\omega-\epsilon_M \pm i\Gamma)^{-1}$.

Substituting the time-dependent solution of $p_1(t)$ and $p_0(t)$
given by \Eq{p1p0t-1} into the current expression \Eq{IL-1},
we obtain the transient current $I^{(1)}_L(t)$ as the sum from the two channels
\bea\label{IL1B-ept}
I^{(1)}_{L}(A)&=&I^{(1)}_{L,st}(A)\, G(t)  \,, \nl
I^{(1)}_{L}(B)&=&\Gamma^{+}_L  [1- G(t)]  \,.
\eea
$I^{(1)}_{L}(A)$ is the transient current associated with
the steady-state-occupation-channel,
while  $I^{(1)}_{L}(B)$ is the current associated with
the empty-occupation-channel.
Remarkably, we stress that here
$I^{(1)}_{L,st}(A)=I^{(1)}_{L,st}(A_1)+I^{(1)}_{L,st}(A_2)+I^{(1)}_{L,st}(A_3)$,
which allows us to obtain the individual transient components.

Let us now consider further the current $I^{(2)}_L(t)$, i.e.,
the current contribution of the ``second" electron in the Andreev process.
Similar analysis can be performed along the line for the current $I^{(1)}_L(t)$.
First, for the steady-state current, we have
\bea\label{IL-e2-1}
I^{(2)}_{L,st}(A_1)&=& \frac{e}{\hbar} \,
\left(\widetilde{\Gamma}^+_L\Gamma^+_L  - \widetilde{\Gamma}^-_L\Gamma^-_L \right)/2\Gamma  \,,  \nl
I^{(2)}_{L,st}(A_2)&=& \frac{e}{\hbar} \,
\left(\widetilde{\Gamma}^+_L\Gamma^+_R  - \widetilde{\Gamma}^-_L\Gamma^-_R  \right)/2\Gamma \,,  \nl
I^{(2)}_{L,st}(A_3)&=& \frac{e}{\hbar} \,
\left(\widetilde{\Gamma}^+_L \widetilde{\Gamma}^-_R -
\widetilde{\Gamma}^-_L \widetilde{\Gamma}^+_R \right)/2\Gamma    \,.
\eea
We notice that $I^{(2)}_{L,st}(A_1)=I^{(1)}_{L,st}(A_1)$, which simply means
the equal contribution of the two electrons
involved in the local AR process, to the left-lead current.
The other two components can be reexpressed as \cite{Li20b}
\bea\label{IL-e2-2}
I^{(2)}_{L,st}(A_2)&=&
\frac{e}{h}\int^{\mu_R}_{-\mu_L} d\omega\, {\cal T}^{eh}_{RLA}(\omega) \,, \nl
I^{(2)}_{L,st}(A_3)&=&
\frac{e}{h}\int^{-\mu_R}_{-\mu_L} d\omega\, {\cal T}^{hh}_{RL}(\omega)  \,.
\eea
It is clear that $I^{(2)}_{L,st}(A_2)$ is the crossed AR process current
associated with a right-lead electron and a left-lead hole,
while $I^{(2)}_{L,st}(A_3)$ is the current owing to
hole-hole tunneling from the right lead to the left one.
Similar as \Eq{T-coef-1}, here the two transport coefficients are obtained as
${\cal T}^{eh}_{RLA}(\omega) = \Gamma^e_R\Gamma^h_L / |Z|^2 $
and ${\cal T}^{hh}_{RL}(\omega) = \Gamma^h_R\Gamma^h_L / |Z|^2$.

Next, the transient current $I^{(2)}_{L}(t)$
can be similarly obtained and be separated into two parts
\bea\label{IL2B-ept}
I^{(2)}_{L}(A)&=&I^{(2)}_{L,st}(A)\, G(t)  \,, \nl
I^{(2)}_{L}(B)&=& -\widetilde{\Gamma}^{-}_L  [1- G(t)]  \,.
\eea
Here the steady-state current
$I^{(2)}_{L,st}(A)$
is the sum of the three components given above, i.e.,
$I^{(2)}_{L,st}(A)=I^{(2)}_{L,st}(A_1)+I^{(2)}_{L,st}(A_2)+I^{(2)}_{L,st}(A_3)$.
$I^{(2)}_{L}(B)$ is the transient current
owing to the gradual decay of the  empty-occupation-channel,
see Fig.\ 2(a) for a schematic illustration.

\begin{figure}[h]
\includegraphics[scale=0.9]{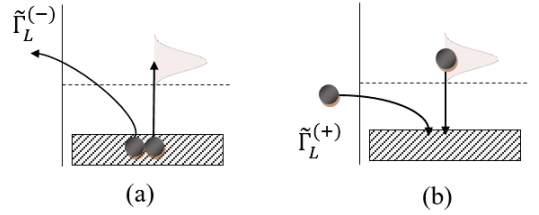}
\caption{
Schematic diagram for transient contribution
to the branch circuit current $\widetilde{I}_L$,
owing to the decay of the initial-occupation-channels.
(a)
The process contributing to $I^{(2)}_L(B)$ given by \Eq{IL2B-ept}
for the unoccupied initial condition, while
$I^{(1)}_L(B)$ given by \Eq{IL1B-ept} is not relevant to the branch circuit current.
(b)
The process contributing to $I^{(2)}_L(B)$ given by \Eq{IL12B-ocp}
for the occupied initial condition, while
$I^{(1)}_L(B)$ is similarly not relevant to the branch circuit current.   }
\end{figure}

\subsection{Occupied Initial Condition}

In this subsection we briefly show the results for the initial condition of
{\it occupied state} of the $f$ particle ($n_f=1$).
The time-dependent solution of the master equation reads
\bea
p_1 &=& \frac{r_1}{2\Gamma} G(t) + [1-G(t)]  \,, \nl
p_0 &=& \frac{r_2}{2\Gamma} G(t)   \,.
\eea
This solution holds also the {\it two-channel} interpretation,
i.e., the gradual formation of the {\it steady-state-occupation-channel}
associated with $G(t)$ (from 0 to 1),
and the gradual decay of the {\it initially-occupied-channel}
described by $1-G(t)$ (from 1 to 0).
Therefore, under the occupied initial condition,
the components of transient current
$I^{(1)}_{L}(A)$ and $I^{(2)}_{L}(A)$
associated with the formation of the steady-state-occupation-channel
are the same as those under the empty initial condition.
However, the transient components $I^{(1)}_{L}(B)$ and $I^{(2)}_{L}(B)$,
owing to the decay of the initially-occupied-channel,
have different results, which read
\bea\label{IL12B-ocp}
I^{(1)}_{L}(B) &=& - \Gamma^{-}_L  [1- G(t)] \,,  \nl
I^{(2)}_{L}(B) &=& \widetilde{\Gamma}^{+}_L  [1- G(t)] \,.
\eea
In Fig.\ 2(b), we illustrate the process of $I^{(2)}_{L}(B)$
for the occupied initial condition.

\section{Cross Correlation of Andreev Current Fluctuations}

As pointed out at the beginning of the previous section,
within the formulation of rate equation (under the Markovian approximation),
the total current, either $I_L$ or $I_R$, is not time dependent
even in the transient process of the occupation probability $p_1(t)$.
However, as analyzed later in detail, we know that
the individual components of the current
are time dependent and can show transient behaviors.
Therefore, we may consider the cross correlation between the currents
$\widetilde{I}_L$ and $\widetilde{I}_R$,
by deducting, from the total currents $I_L$ and $I_R$, the components of
the electron-electron and hole-hole transmission between the left and right leads.
More explicitly, we introduce
$\widetilde{I}_{L}(t)=I_{L}-I^{(1)}_{L}(A_3)-I^{(2)}_{L}(A_3)$
and $\widetilde{I}_{R}(t)=I_{R}-I^{(1)}_{R}(A_3)-I^{(2)}_{R}(A_3)$,
and consider their correlation.
It is clear that, as shown schematically in Fig.\ 1,
$\widetilde{I}_{L}$ and $\widetilde{I}_{R}$ are, respectively,
the branch circuit currents flowing back to the left and right leads
from the superconductor, being associated with the Andreev process.
In this work, we may thus term
$\widetilde{I}_{L}$ and $\widetilde{I}_{R}$ {\it Andreev currents}.

Following the standard formulation,
let us consider the correlator of current fluctuations,
$S_{LR}(t) = \la  \delta\widetilde{I}_L(t) \delta\widetilde{I}_R(0) \ra_{\bar{\rho}}$,
where $\delta\widetilde{I}_{\alpha}(t)=\widetilde{I}_{\alpha}(t)
- \la \widetilde{I}_{\alpha} \ra_{\bar{\rho}} $.
$\la \cdots \ra_{\bar{\rho}}$ means making average over the steady state $\bar{\rho}$.
Actually, we know that
$S_{LR}(t) = \la \widetilde{I}_L(t)\widetilde{I}_R(0) \ra_{\bar{\rho}}
- \la \widetilde{I}_L\ra_{\bar{\rho}}\la \widetilde{I}_R\ra_{\bar{\rho}} $.
We thus focus on calculating the correlator
$\widetilde{S}_{LR}(t) = \la \widetilde{I}_L(t)\widetilde{I}_R(0) \ra_{\bar{\rho}}$.
Applying the {\it quantum jump} approach \cite{WM09}, we know that
after measuring the current $\widetilde{I}_R$ at the first time ($t=0$),
the steady state is changed to be
\bea
\widetilde{\rho}(0)= a |0\ra \la 0|+ b |1\ra \la 1| \,,
\eea
while the two coefficients read as
\bea
a &=& \left[ \widetilde{\Gamma}^{+}_R
(\Gamma^{+}_R + \Gamma^{+}_L) - \Gamma^{-}_R
(\widetilde{\Gamma}^{-}_R + \widetilde{\Gamma}^{-}_L ) \right]/2\Gamma \,,  \nl
b &=& \left[{\Gamma}^{+}_R
(\widetilde{\Gamma}^{+}_R + \widetilde{\Gamma}^{+}_L)
- \widetilde{\Gamma}^{-}_R (\Gamma^{-}_R + \Gamma^{-}_L )  \right]/2\Gamma \,.
\eea
Now, $\widetilde{\rho}(0)$ is no longer the steady state.
Its evolution before the second measurement of $\widetilde{I}_L$
at time $t$ is formally given by
\bea
\widetilde{\rho}(t) = e^{{\cal L}t} (a |0\ra \la 0|+ b |1\ra \la 1|) \,,
\eea
where the propagator (in terms of the Liouvillian superoperator)
stands for the evolution governed by the MME.
After knowing this result, the correlator
can be computed straightforwardly through
$\widetilde{S}_{LR}(t) = \la \widetilde{I}_L\ra_{\widetilde{\rho}(t)}$,
which contains two parts,
$\widetilde{S}_{LR}(t)=\widetilde{S}^{(I)}_{LR}(t)+\widetilde{S}^{(II)}_{LR}(t)$.
The first part is contributed by the disturbed evolution associated with
{\it the formation of the steady-state-occupation-channel},
which is given by
\bea
\widetilde{S}^{(I)}_{LR}(t) = \frac{e}{\hbar} (a+b)
\left( \widetilde{I}^{(1)}_{L,st} + \widetilde{I}^{(2)}_{L,st} \right) G(t) \,,
\eea
where the two components of the steady-state current
flowing back from the superconductor to the left lead, i.e.,
$\widetilde{I}^{(1)}_{L,st} = I^{(1)}_{L,st}(A_1) + I^{(1)}_{L,st}(A_2)$
and $\widetilde{I}^{(2)}_{L,st} = I^{(2)}_{L,st}(A_1) + I^{(2)}_{L,st}(A_2)$,
read as
\bea
\widetilde{I}^{(1)}_{L,st} &=& \frac{e}{\hbar} \left[\Gamma^{+}_L
(\widetilde{\Gamma}^{+}_L + \widetilde{\Gamma}^{+}_R)
- \Gamma^{-}_L
(\widetilde{\Gamma}^{-}_L + \widetilde{\Gamma}^{-}_R)  \right] /2\Gamma \,,  \nl
\widetilde{I}^{(2)}_{L,st} &=& \frac{e}{\hbar} \left[\widetilde{\Gamma}^{+}_L
(\Gamma^{+}_L + \Gamma^{+}_R)
- \widetilde{\Gamma}^{-}_L
(\Gamma^{-}_L + \Gamma^{-}_R)  \right] /2\Gamma   \,.
\eea
Let us denote
$\widetilde{I}^{(1)}_{L,st}+\widetilde{I}^{(2)}_{L,st}
=\la\widetilde{I}_L \ra_{\bar{\rho}}$,
One can also easily check that
$\frac{e}{\hbar} (a+b)=\la\widetilde{I}_R \ra_{\bar{\rho}}$.
Thus, we arrive at a compact expression
\bea
\widetilde{S}^{(I)}_{LR}(t)
= \la\widetilde{I}_L \ra_{\bar{\rho}}\la\widetilde{I}_R \ra_{\bar{\rho}} G(t) \,.
\eea
Moreover, the second part of the correlator, $\widetilde{S}^{(II)}_{LR}(t)$,
which is associated with {\it the decay of the initial-occupation-channels}
(a statistical mixture of the occupied
and unoccupied states of the $f$ particle, see Fig.\ 2), can be obtained as
\bea
\widetilde{S}^{(II)}_{LR}(t)
= \left(\frac{e}{\hbar} \right)^2
 \left(b\widetilde{\Gamma}^{+}_L - a\widetilde{\Gamma}^{-}_L\right) [1-G(t)] \,.
\eea
Finally, we obtain the current-fluctuation-correlator as
\bea
&& S_{LR}(t)= \widetilde{S}_{LR}(t)
- \la\widetilde{I}_L \ra_{\bar{\rho}}\la\widetilde{I}_R \ra_{\bar{\rho}}  \nl
&& = \left[ \left(\frac{e}{\hbar}\right)^2
\left(b\widetilde{\Gamma}^{+}_L - a\widetilde{\Gamma}^{-}_L\right)
-\la\widetilde{I}_L \ra_{\bar{\rho}}\la\widetilde{I}_R \ra_{\bar{\rho}} \right] e^{-2\Gamma t} \nl
&&\equiv ~ C_{LR}\,\, e^{-2\Gamma t}  \,.
\eea
Here, we introduced the cross-correlation-factor $C_{LR}$
which characterizes the essential correlation property,
by noting that the time dependent behavior is relatively simple,
which simply leads to a Lorentzian lineshape
$\sim (\omega^2+4\Gamma^2)^{-1}$ in frequency domain.
The most important point we would like to emphasize here is that
$C_{LR}\neq 0$, even when the Majorana coupling energy $\epsilon_M\to 0$.

We notice that in Refs.\ \cite{Dem07,Bee08,Zoch13} it was found that
the cross correlation between the total currents in the different leads
vanishes at the limit $\epsilon_M\to 0$.
Here, despite
using the total-occupation-probability-description based master equation approach,
we have successfully isolated the component currents
(i.e., the branch circuit currents) from the total ones,
and found nonzero cross correlation even at the limit $\epsilon_M\to 0$.
In the occupation-number-state representation, the picture
of nonzero cross correlation appears clear.
Owing to the nonlocal nature of the MZMs associated $f$ quasiparticle,
it is natural to expect that the disturbance at one side
should influence the electron-hole excitation at the other side,
leading thus to the nonlocal crossed AR process.
In certain sense, the nonlocal $f$ particle is quite similar as
the single-level electron in a quantum dot which, obviously,
can correlate the currents in the two leads of transport.
In the Majorana case, the left-side sum of the crossed AR and local AR currents,
i.e., the branch circuit current considered in this work,
is similarly correlated with the branch circuit current on the right side,
through the nonlocal $f$ particle.

\begin{figure}[h]
\includegraphics[scale=0.25]{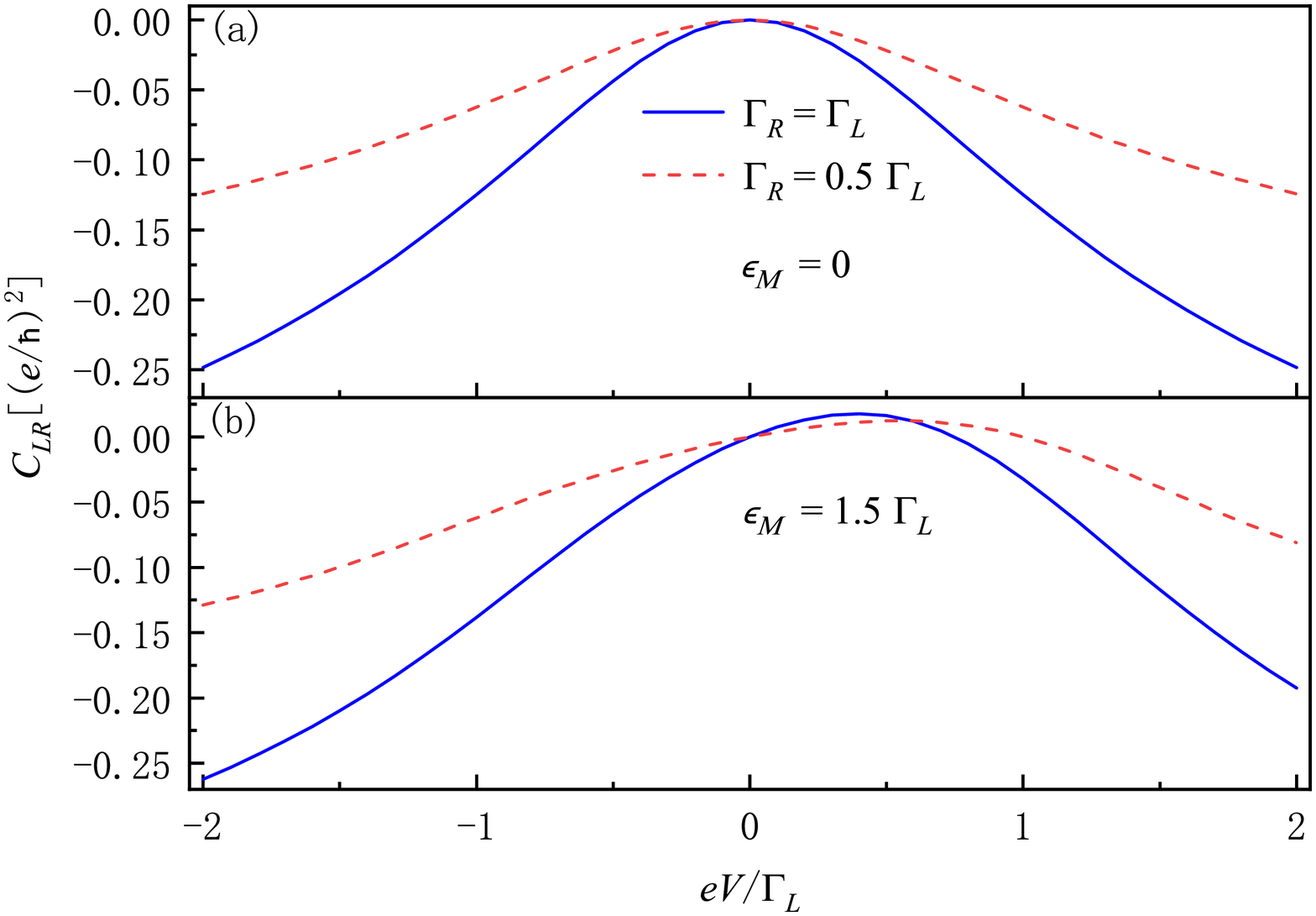}
\caption{
Cross correlation factor $C_{LR}$ as a function of the equally biased voltage
considered in Ref.\ \cite{Bee08}, i.e., $\mu_L=\mu_R=eV$
with respect to the chemical potential of the superconductor ($\epsilon_F=0$).
In (a) and (b), results for the ideal ($\epsilon_M=0$) and nonideal ($\epsilon_M\neq0$)
Majorana cases are shown, together with the symmetric ($\Gamma_L=\Gamma_R$)
and asymmetric ($\Gamma_L\neq\Gamma_R$) coupling to the leads. }
\end{figure}

\begin{figure}[h]
\includegraphics[scale=0.25]{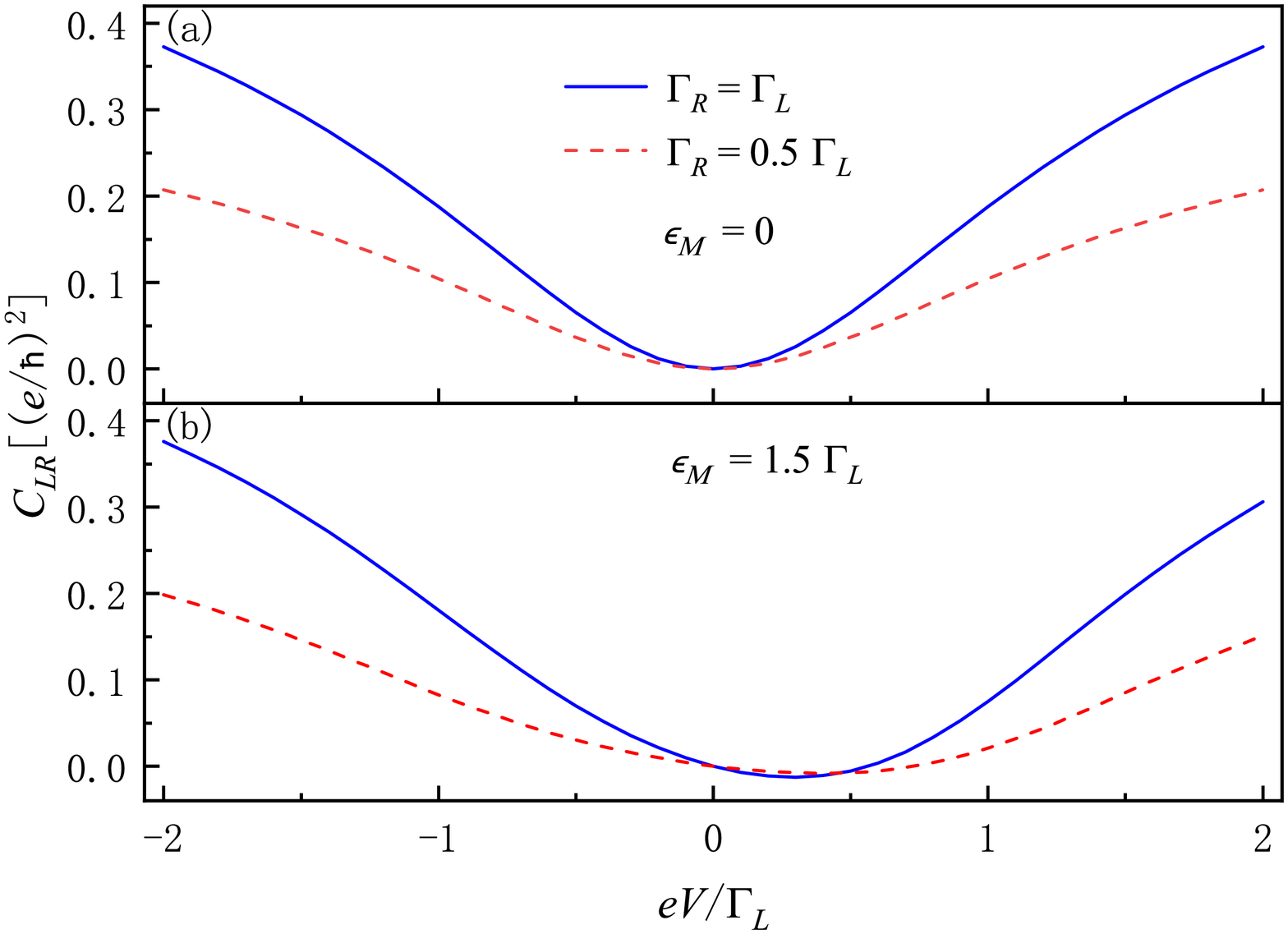}
\caption{
Cross correlation factor $C_{LR}$ as a function of the
anti-symmetrically biased voltage, i.e., $\mu_L=-\mu_R=eV$
with respect to the chemical potential of the superconductor.
Other parametric explanation is the same as in Fig.\ 3. }
\end{figure}

We show in Fig.\ 3 the numerical results of the cross correlation factor $C_{LR}$
as a function of the equally biased voltage
considered in Ref.\ \cite{Bee08}, i.e., $\mu_L=\mu_R=eV$
with respect to the chemical potential of the superconductor ($\epsilon_F=0$).
Indeed, even in the case of ideally uncoupled MZMs (with $\epsilon_M=0$),
the cross correlation is found nonzero.
More specifically, we find $C_{LR}$ negative and can understand it as follows.
Let us imagine a ``quantum jump" associated with confirming
the formation of a Cooper pair in the superconductor.
The observation of this specific event allows us to define
the current $\widetilde{I}_R$
which must be larger the average current $\la \widetilde{I}_R\ra_{\bar{\rho}}$.
We then have $\delta\widetilde{I}_R>0$ at the right side.
After observing this event, the state would jump to $\widetilde{\rho}(0)=|0\ra \la 0|$,
owing to the fact that the formation of a Cooper pair must annihilate the $f$ quasiparticle.
Using this disturbed new state, we calculate the left side current and obtain
\bea
\la \widetilde{I}_L\ra_{\widetilde{\rho}(0)}=-\frac{e}{\hbar}\, \widetilde{\Gamma}^-_L \,.
\eea
This current, generated by the inverse AR process, is negative.
We then have $\delta\widetilde{I}_L<0$ and
arrive at the negative correlation as shown in Fig.\ 3(a),
by noting that the steady state current
$\la \widetilde{I}_L\ra_{\bar{\rho}}$ is positive.

For the bias voltage dependence,
the reason of the symmetric feature is relatively simple.
That is, changing the bias voltage from positive to negative,
the Andreev process is reversed.
Then, the directions of both currents are reversed
and the correlation of the current fluctuations keeps the sign unchanged.
In Fig.\ 3, the effect of asymmetry of coupling to the leads is also shown,
for both the ideal ($\epsilon_M=0$) and nonideal ($\epsilon_M\neq 0$) cases.

In Fig.\ 4, we show the results for the setup of
anti-symmetrically biased voltage, i.e., $\mu_L=-\mu_R=eV$
with respect to the chemical potential of the superconductor.
Here, it is interesting to find that the correlation is {\it positive},
rather than being {\it negative} as seen in Fig.\ 3.
For this bias setup (e.g., $eV>0$), the inverse Andreev process on the right side
leads to a negative current of $\widetilde{I}_R$.
Now, let us consider observing the ``jump" event of splitting a Cooper pair.
This allows us to determine the current which must be larger than the average current,
i.e., $-\widetilde{I}_R > - \la \widetilde{I}_R\ra_{\bar{\rho}}$, thus $\delta\widetilde{I}_R<0$.
After observing this ``jump" event,
the state is updated as $\widetilde{\rho}(0)=|1\ra \la 1|$,
owing to the fact that splitting a Cooper pair must generate the $f$ quasiparticle.
Using this disturbed new state, we calculate the left side current and obtain
\bea
\la \widetilde{I}_L\ra_{\widetilde{\rho}(0)}= \frac{e}{\hbar}\, \widetilde{\Gamma}^+_L \,.
\eea
Moreover, we can straightforwardly examine that
$\frac{e}{\hbar}\, \widetilde{\Gamma}^+_L - \la \widetilde{I}_L\ra_{\bar{\rho}} <0$,
under the bias voltage $eV>0$.
We can thus conclude the positive correlation shown in Fig.\ 4.
Other behaviors, such as the bias voltage and coupling asymmetry dependence,
can be understood similarly as above for the results of Fig.\ 3.

\section{Summary}

Instead of the cross correlation between the total lead currents
in a two-lead (three-terminal) Majorana device, 
we have analyzed the cross correlation of 
the Andreev-process-associated branch circuit currents.
Despite applying a master equation approach, which deals with 
the total occupation probability of the MZMs associated $f$ quasiparticle,
we have been able to successfully extract out 
the component currents from the total ones
and computed the cross correlation by means of the quantum jump technique. 
Importantly, we found that the cross correlation does not vanish 
even when the Majorana coupling energy $\epsilon_M\to 0$, 
owing to the nonlocality nature of the MZMs.
For different setup of bias voltage,
we also found intriguing results of both negative and positive correlations
and carried out simple physical understanding.
The proposed cross correlation between the branch circuit currents 
are experimentally accessible in practice.
Combining with the evidence of the zero-bias-peak of conductance,
the nonlocal cross correlation predicted in this work
can help to definitely confirm the nonlocal MZMs.

%\clearpage
\vspace{0.5cm}
{\flushleft\it Acknowledgements.}---
This work was supported by the
National Key Research and Development Program of China
(No.\ 2017YFA0303304) and the NNSF of China (Nos.\ 11675016, 11974011 \& 61905174).

\end{CJK*}

\end{document}